\documentclass[preprint,12pt]{elsarticle}

\usepackage[margin=2cm]{geometry}

\usepackage[utf8]{inputenc}
\usepackage[T1]{fontenc}
\usepackage{mathptmx}
\usepackage{amsmath,amssymb}
\usepackage{graphicx}
\usepackage{dcolumn}
\usepackage{bm}
\usepackage{physics}
\usepackage{siunitx}
\usepackage{url}
\usepackage[hidelinks]{hyperref}
\usepackage{lipsum}
\makeatletter
\def\ps@pprintTitle{%
 \let\@oddhead\@empty
 \let\@evenhead\@empty
 \def\@oddfoot{}%
 \let\@evenfoot\@oddfoot}
\makeatother

\begin{document}

\begin{frontmatter}

\title{Strain control of magneto-optical properties of ultra-thin Ni-Mn-Ga films grown by magnetron sputtering}

\author[fzu,cuni]{Mat\v{e}j Make\v{s}\corref{cor1}}
\ead{makes@fzu.cz}
\author[fzu]{Petr Ve\v{r}t\'{a}t}
\author[fzu]{Michal Rame\v{s}}
\author[cuni]{Jakub Z\'{a}zvorka}
\author[fzu]{Franti\v{s}ek M\'{a}ca}
\author[fzu,cvut]{Jan Zemen}
\author[fzu]{Oleg Heczko}
\author[cuni]{Martin Veis}

\cortext[cor1]{Corresponding author}

\address[fzu]{FZU -- Institute of Physics of the Czech Academy of Sciences, Na Slovance 1999/2, 18200 Prague 8, Czechia}
\address[cuni]{Faculty of Mathematics and Physics, Charles University, Ke Karlovu 3, 12116 Prague 2, Czechia}
\address[cvut]{Faculty of Electrical Engineering, Czech Technical University, Technick\'{a} 2, Prague 6, 166 00, Czech Republic}

\begin{abstract}
Ultra-thin epitaxial Ni-Mn-Ga films, \SIrange{10}{15}{\nm} thick, were grown by magnetron sputtering on MgO(001), either directly on the substrate or on a Cr buffer, and studied by X-ray diffraction, magnetometry, and magneto-optical Kerr effect (MOKE) spectroscopy. The films were austenitic, with the cubic cell tetragonally distorted in opposite directions. The film on bare MgO was stretched in-plane ($c/a=0.989$) and the film on Cr was compressed in-plane ($c/a=1.021$), while their unit-cell volumes differed by less than \SI{0.3}{\percent}. As the thermal expansion mismatch was expected to stretch the films in-plane in both cases, the opposite distortions were attributed to the lattice misfit. The room-temperature magnetization loops of the two films did not show significant differences, but their polar and longitudinal MOKE spectra differed in the amplitude, sign, and energy of the spectral structures over the whole \SIrange{0.76}{5}{\eV} range. Magneto-optical simulations showed that the absence of the Cr buffer accounted only for the stronger low-energy polar MOKE rotation extremum of the film on MgO. We attributed the other changes to the strain-induced change of the electronic structure, consistent with theoretical predictions in which the $d_{x^2-y^2}$ and $d_{3z^2-r^2}$ sub-bands of the minority-spin Ni $e_g$ states exchange their order when the distortion changes sign. In parts of the spectrum, the two films showed Kerr rotation or ellipticity of opposite sign, so the choice of the growth template or a dynamically induced distortion can switch the sign of the magneto-optical response at a given photon energy.
\end{abstract}


\end{frontmatter}


\section{Introduction}

The magnetic shape memory alloy Ni-Mn-Ga is a ferromagnetic Heusler alloy whose cubic L2$_1$ austenite transforms into lower-symmetry martensitic phases. Since its discovery in a Ni$_2$MnGa single crystal~\cite{ullakko1996}, the magnetic-field-induced strain caused by twin-boundary motion has reached, respectively, \SIlist{5.1;9.5;12}{\percent} in bulk single crystals of the 10M, 14M, and NM martensites~\cite{heczko2000,sozinov2002,sozinov2013}. Band-structure calculations attributed the instability of the cubic lattice to a band Jahn--Teller mechanism~\cite{fujii1989}, supported by polarized neutron scattering~\cite{brown1999}. In cubic Ni$_2$MnGa, the states closest to the Fermi level ($E_\text{F}$) form a sharp minority-spin peak of degenerate Ni $e_g$ ($d_{x^2-y^2}$, $d_{3z^2-r^2}$) character, calculated about \SI{0.2}{\eV} below~\cite{barman2005} or directly at $E_\text{F}$~\cite{obata2023} and observed by photoemission 0.3--\SI{0.35}{\eV} below $E_\text{F}$~\cite{opeil2008,kimura2013}.

The total energy of Ni$_2$MnGa depends only weakly on $c/a$, consistent with the wide range of observed tetragonal distortions~\cite{godlevsky2001}. A tetragonal distortion lifts the degeneracy of the Ni $e_g$ peak, and the energy order of the two sub-bands reverses with the sign of the distortion. For $c/a>1$, the $d_{3z^2-r^2}$ states move below and the $d_{x^2-y^2}$ states above $E_\text{F}$~\cite{obata2023}, whereas for $c/a<1$ the $d_{3z^2-r^2}$ peak moves up towards $E_\text{F}$ and the $d_{x^2-y^2}$-dominated peak moves down~\cite{barman2005}. For distortions as small as $|c/a-1|\lesssim2\%$, i.e., smaller than the distortions $c/a=0.97$ and $1.20$ considered in these calculations, the occupation of the split sub-bands depends on whether the cubic peak lies below $E_\text{F}$, as in the generalized gradient approximation (GGA), or at $E_\text{F}$, as in quasiparticle self-consistent $GW$ (QSGW). The effect of such small distortions on the electronic structure has not been tested experimentally.

For integration into microsystems, Ni-Mn-Ga is deposited as thin films, usually hundreds of nanometres or more thick, most commonly on MgO(001)~\cite{thomas2008stress,buschbeck2009,luo2011,ranzieri2013,jakob2013}. 
\citet{thomas2008stress} showed for a \SI{473}{\nm} thick film that its larger thermal expansion coefficient than that of its substrate, together with a lattice misfit~\cite{luo2011} of about 2.5\% produce a biaxial tensile stress of up to \SI{105}{\mega\pascal}, which stretches the austenite in-plane ($c/a<1$) and raises its martensitic transformation temperature by about \SI{63}{\kelvin}. A Cr buffer has low misfit to Ni-Mn-Ga, which enables epitaxial growth without significant interdiffusion and the release of freestanding films by selective etching~\cite{backen2010,eichhorn2011}. Ultra-thin films, in which the substrate constraint acts on the whole film volume, have received much less attention. Films \SIlist{10;20}{\nm} thick on MgO(001) remain austenitic over the whole investigated temperature range~\cite{ranzieri2013}, as do Cr-buffered films up to \SI{38}{\nm} thick~\cite{makes2026}. In such films the interface may also contribute, as Cr monolayers on Ni$_2$MnGa(001) were predicted to carry strongly enhanced magnetic moments~\cite{bhattacharya2021} and interfacial spin--orbit interaction modifies the polar Kerr effect of ultra-thin Fe/GaAs(001) layers~\cite{buchner2016}.

Magneto-optical Kerr effect (MOKE) spectroscopy is well suited to probe such changes, as it depends crucially on band-structure details such as spin--orbit coupling and exchange splitting~\cite{oppeneer1992}. In a Ni-Mn-Ga single crystal, the polar Kerr rotation reversed its sign between 2 and \SI{3}{\eV} upon the martensitic transformation, without a shift of the main spectral bands~\cite{veis2014,beran2015}. In Cr-buffered films, the polar MOKE spectra evolved with thickness~\cite{makes2026}, but the tetragonal distortion of the austenite ($c/a=\numrange[range-phrase=-]{1.004}{1.02}$) changed together with the thickness~\cite{makes2026}. The effect of a small strain-induced distortion, and in particular of its sign, on the magneto-optical spectra has thus not been isolated, nor have interface optical effects been separated from strain effects. Here we compare \SIlist{10;15}{\nm} thick films grown under identical conditions directly on MgO(001) and on a Cr buffer, with opposite distortions of the austenite, $c/a=0.989$ and $1.021$. As the optical penetration depth (below \SI{25}{\nm}~\cite{makes2026}) is comparable to the film thickness, MOKE probes the whole strained film including its interface. We show that these small distortions produce markedly different polar and longitudinal MOKE spectra and use optical simulations to separate the optical effect of the buffer from the change of the electronic structure.

\section{Experimental}

A series of ultra-thin epitaxial Ni-Mn-Ga films was prepared using magnetron sputtering at \SI{550}{\celsius} deposition temperature and \SI{e-9}{\Pa} vacuum on MgO(100) substrate.
Two sets of films with different nominal thicknesses (\SIlist{10;15}{\nm}) were prepared with and without \SI{20}{\nm} Cr buffer and a reference \SI{20}{\nm} Cr buffer without Ni-Mn-Ga film.

X-ray diffraction (XRD) and X-ray reflectivity (XRR) measurements were carried out using a Rigaku SmartLab diffractometer with a Cu anode ($\lambda_{\text{K}\alpha 1} = \SI{0.15406}{\nm}$, $\lambda_{\text{K}\alpha 2} = \SI{0.15444}{\nm}$) in parallel-beam geometry. The lattice parameters were established from symmetric and asymmetric $\theta$-$2\theta$ scans of strong fundamental reflections, and the XRR data were analyzed using GenX 3 python package developed by \citet{glavic2022}.

Magnetization loops were measured at room temperature using Quantum Design PPMS up to \SI{9}{\tesla}. The linear diamagnetic contribution from the MgO substrate was subtracted from the data.

The polar magneto-optical Kerr effect (PMOKE) was measured with a custom-built rotating polarizer ellipsometer (RPE) at near-normal incidence in $\pm\SI{1}{\tesla}$ applied magnetic field from ultra-violet (UV) to infra-red (IR) range, i.e., in the \SIrange[range-phrase=--]{0.76}{5}{\eV} spectral range. We used Ocean Optics QEPro spectrometer for the ultra-violet (UV) and visible (VIS) regions and spectrometer Ocean Optics Flame NIR for the near-infra-red (NIR) region. To eliminate magnetization-even effects, spectra were averaged over both magnetic field polarities and 10 magnetizing cycles to obtain precision of up to a few millidegrees.

\section{Results and discussion}

Nominal compositions, thicknesses, and lattice parameters of the studied films are listed in Table~\ref{tab:sample_info}. All films grew epitaxially in the austenitic phase with tetragonally distorted cubic lattice. Film lattice grew rotated by \ang{45} in-plane with respect to MgO and parallel to Cr lattice, which is itself slightly tetragonally distorted ($c_\text{Cr}/a_\text{Cr}=0.997$). Lattice parameters were determined for the nominally \SI{15}{\nm} thick films, which show that the two templates distort the cubic cell in opposite directions. On bare MgO(001), the film is stretched in-plane and shortened out-of-plane ($c/a=0.989$) without fully adopting the substrate lattice with residual misfit $(a-\sqrt{2}a_\text{MgO})/\sqrt{2}a_\text{MgO}=\SI{-1.7}{\percent}$. On the Cr buffer, the film has a smaller misfit $(a-2a_\text{Cr})/2a_\text{Cr}=\SI{0.2}{\percent}$, is compressed in-plane, and elongated out-of-plane ($c/a=1.021$). Remarkably, the unit-cell volumes $ca^2$ of both films (\SIlist{198.3;198.8}{\angstrom\cubed}) differ by less than \SI{0.3}{\percent}, corresponding to the same equivalent cubic lattice parameter of \SI{0.583}{\nm}. The two films thus differ essentially only in the sign of the tetragonal distortion, $c/a-1=\SI{-1.1}{\percent}$ and $+\SI{2.1}{\percent}$.

The distortions are set by the template during growth at \SI{550}{\celsius} and modified by the subsequent cooling. As described by \citet{chernenko2006thickness}, a film clamped to a much thicker substrate acquires on cooling an in-plane strain $(\alpha_\text{f}-\alpha_\text{s})\Delta T$, which is tensile when the linear thermal expansion coefficient of the film, $\alpha_\text{f}$, exceeds that of the substrate, $\alpha_\text{s}$. The coefficient of Ni-Mn-Ga films~\cite{chernenko2006thickness}, \SI{15e-6}{\per\kelvin}, exceeds those of MgO~\cite{touloukian1977}, \SI{12e-6}{\per\kelvin}, and of Cr near room temperature~\cite{koumelis1973}, \SI{5.1e-6}{\per\kelvin} below and \SI{8.6e-6}{\per\kelvin} above its N\'eel temperature (found at \SI{43.3}{\celsius} in that study). Cooling to room temperature ($\Delta T\approx\SI{525}{\kelvin}$) thus stretches a Ni-Mn-Ga film clamped to MgO in-plane by up to about \SI{0.16}{\percent}. The Cr buffer, which contracts less than MgO, is thermally compressed in-plane by about \SI{0.2}{\percent}, but its in-plane lattice parameter still exceeds the out-of-plane one ($c_\text{Cr}/a_\text{Cr}=0.997$--$0.998$), so the misfit to the larger MgO lattice outweighs the thermal contribution. Since the \SI{22}{\nm} thick buffer is expected to follow the thermal contraction of the much thicker MgO substrate, the thermal strain of the Ni-Mn-Ga film is tensile on both templates, and the opposite sign of the distortion must originate in the lattice misfit. On Cr, the doubled in-plane lattice parameter of the buffer, $2a_\text{Cr}=\SI{0.5784}{\nm}$, is smaller than the equivalent cubic parameter of the film, \SI{0.583}{\nm}, and the film is compressed in-plane by about \SI{0.6}{\percent}. On MgO, $\sqrt{2}a_\text{MgO}=\SI{0.5955}{\nm}$ is larger, and the film, with most of the misfit relaxed, is stretched in-plane by about \SI{0.4}{\percent}. The thermal strain accounts for less than half of this stretch, similar to the observation of Thomas et al.~\cite{thomas2008stress} that the measured tensile strain is about three times the calculated thermal strain.

\begin{table*}[t]
    \centering
    \caption{Nominal composition and thickness (in nm—found in brackets following composition
of each layer) of Ni-Mn-Ga epitaxial films deposited on MgO substrate were derived,
respectively, from sputtering target composition and deposition times. Thicknesses $d$, $d_\text{Cr}$ were determined by XRR, tetragonal lattice parameters $c$, $a$, $c_\text{Cr}$, $a_\text{Cr}$, and cubic $a_\text{MgO}$ were determined by XRD.}
    \label{tab:sample_info}
    \footnotesize\setlength{\tabcolsep}{2pt}
    \begin{tabular}{lcccccccc}
    \hline\hline 
    Nominal composition &
    $d$ (nm)&
    $d_\text{Cr}$ (nm) &
    $c$ (nm)&
    $a$ (nm)&
    $c/a$ &
    $c_\text{Cr}$ (nm)&
    $a_\text{Cr}$ (nm)&
    $c_\text{Cr}/a_\text{Cr}$ \\
    \hline

    Ni$_{48}$Mn$_{30}$Ga$_{22}(10)$ & 
    8.7(2) &
    -- &
    -- &
    -- &
    -- &
    --&
    --&
    -- \\

    Cr(20)/Ni$_{48}$Mn$_{30}$Ga$_{22}(10)$ &
    9.0(2) &
    22.3(2) &
    -- &
    -- &
    -- &
    --&
    --&
    -- \\
    
    Ni$_{48}$Mn$_{30}$Ga$_{22}(15)$ & 
    12.3(2) &
    -- &
    0.5789 &
    0.5853 &
    0.989 &
    --&
    --&
    -- \\
  
    Cr(20)/Ni$_{48}$Mn$_{30}$Ga$_{22}(15)$ & 
    13.2(2) &
    22.3(2) &
    0.5917 &
    0.5797 &
    1.021 &
    0.2884 &
    0.2892&
    0.997 \\

    Cr(20) & 
    -- &
    25.1(3) &
    -- &
    -- &
    -- &
    0.2885 &
    0.2891 &
    0.998 \\
    \hline
    \multicolumn{9}{l}{\textit{Note:} $\sqrt{2}a_\text{MgO}=\SI{0.5955}{\nm}$}\\
    \hline\hline
    \end{tabular}
\end{table*}

\begin{figure}[h]
     \centering
     \includegraphics[width=0.7\linewidth]{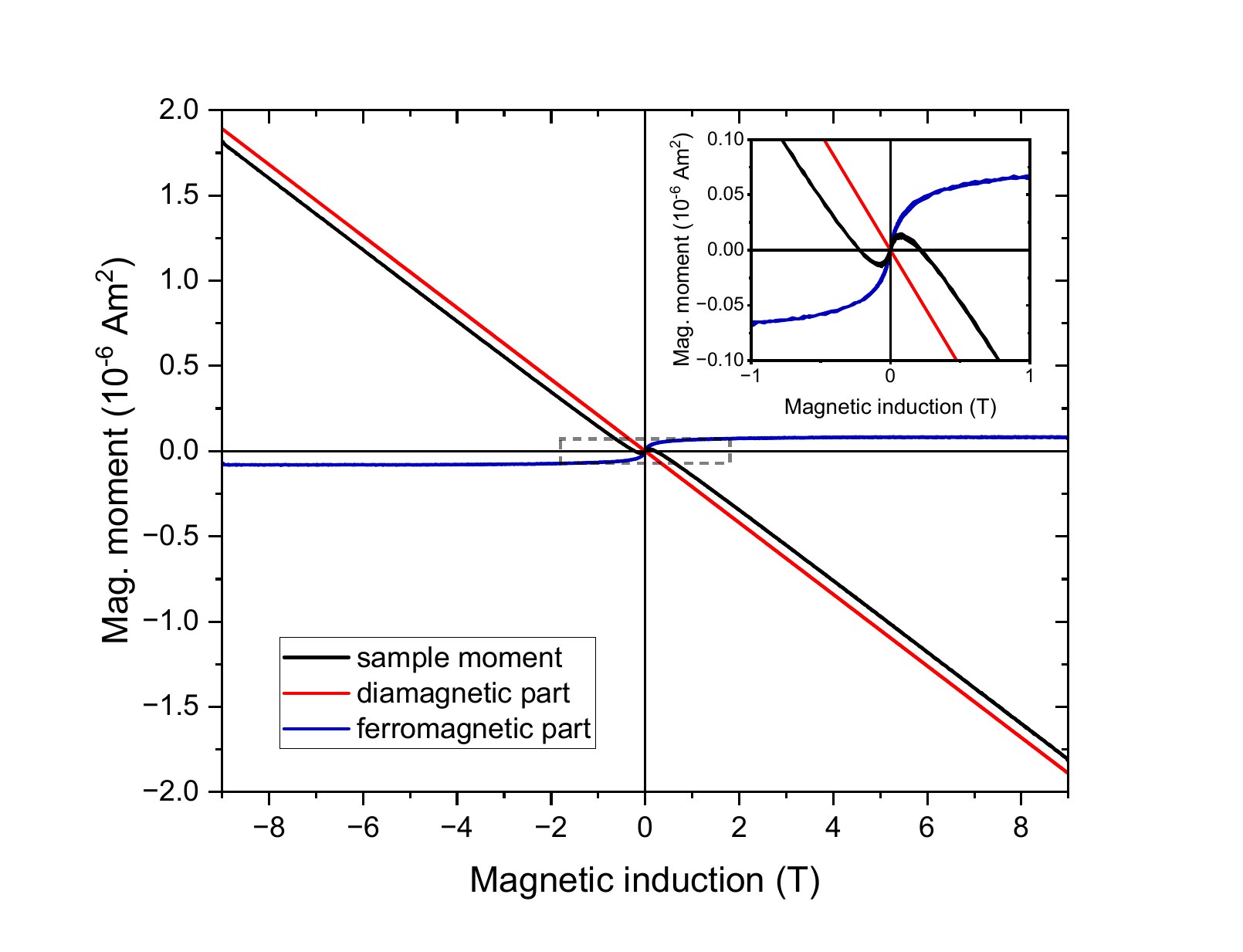}
     \caption{Room-temperature magnetization loop of \SI{15}{\nm}-thick Ni-Mn-Ga film with Cr-buffer (black) decomposed into diamagnetic (red) and ferromagnetic (blue) contributions.}
     \label{fig:300K_PPMS_J170_raw}
 \end{figure}

Comparison of magnetization loops for films with and without Cr buffer did not reveal expected reorientation of easy axis due to the substrate constraint. As seen from Fig.~\ref{fig:300K_PPMS_J170_raw}, the magnetization signal is mainly dominated by diamagnetic contribution of MgO substrate. After the subtraction of this contribution the loops of both samples (with and without Cr layer) exhibit the same magnetization behavior within the given experimental accuracy.

\begin{figure}
    \centering
    \includegraphics[width=0.7\linewidth]{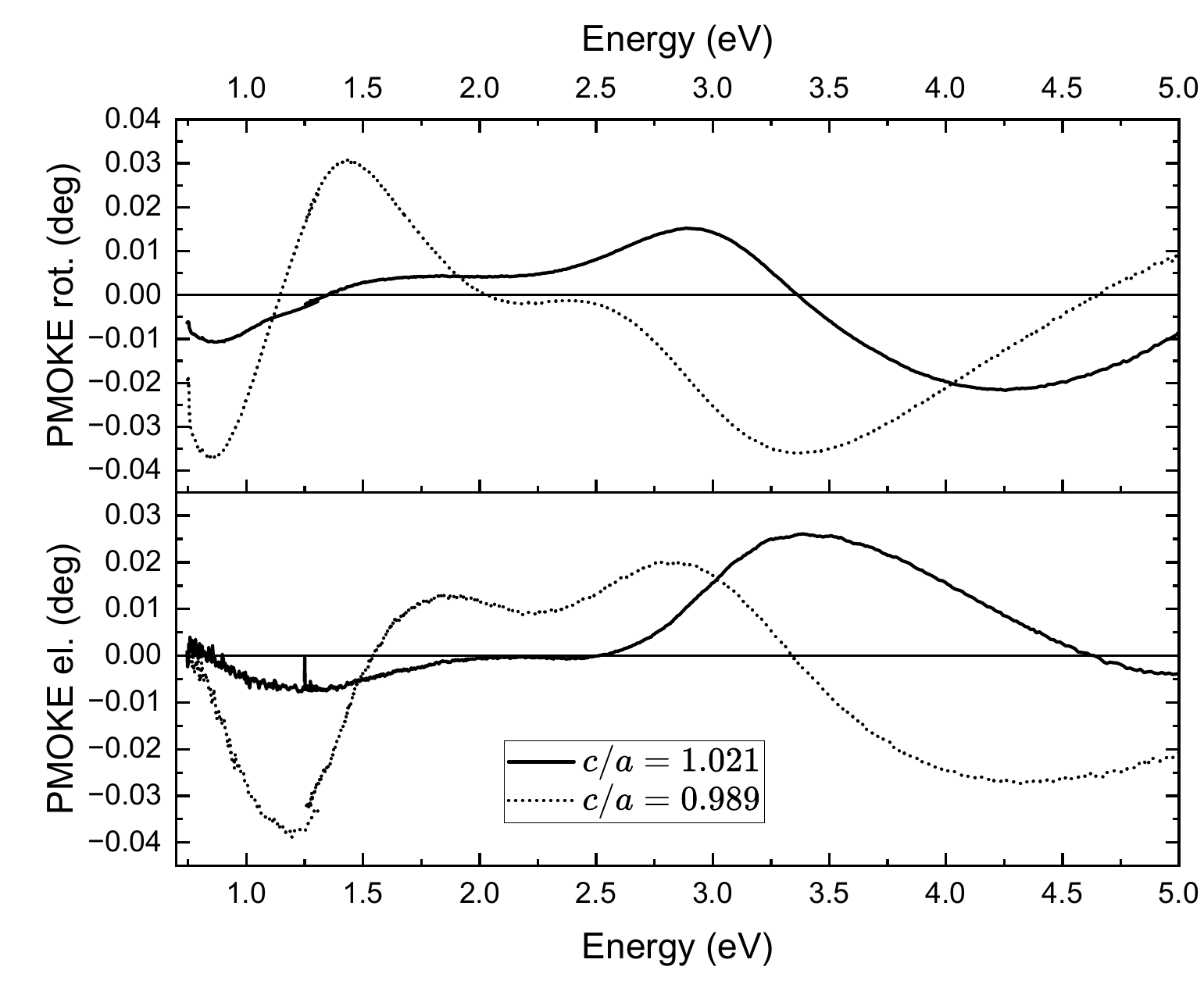}
    \caption{Strain controlled normal incidence PMOKE rotation (top) and ellipticity (bottom) spectra in 15-nm-thick Ni-Mn-Ga films with (solid black lines) and without (dotted black lines) Cr buffer grown with elongated or shortened the originally cubic unit cell in the out-of-plane direction ($c$-axis) as evident from the $c/a$ ratio.}
    \label{fig:Cr_infl_PMOKE_15nm}
\end{figure}

Although the magnetization response was effectively indistinguishable in the films, the PMOKE spectra differ strongly (Fig.~\ref{fig:Cr_infl_PMOKE_15nm}). While both PMOKE rotation spectra share a negative extremum at \SI{0.85}{\eV}, in most of the VIS-to-UV region,
they have different spectral structures present. For $c/a=0.989$, the PMOKE rotation is dominated by a strong positive peak at \SI{1.45}{\eV}, followed by a weak feature near \SI{2.4}{\eV} and a negative extremum at \SI{3.4}{\eV}. For $c/a=1.021$, the \SI{1.45}{\eV} peak collapses into a plateau between \SIlist{1.5;2.3}{\eV}, whereas the higher-energy positive extremum is enhanced and found \SI{0.5}{\eV} higher, at \SI{2.9}{\eV}, and the UV extremum moves to \SI{4.25}{\eV}. The PMOKE ellipticity spectra show similar trends as they are bound by Kramers-Kronig relations~\cite{levallois2015}. The nominally \SIlist{10;15}{\nm} thick films differed only in overall amplitude, which can be attributed to surface quality and probed volume, not in spectral shape.

\begin{figure}
    \centering
    \includegraphics[width=0.7\linewidth]{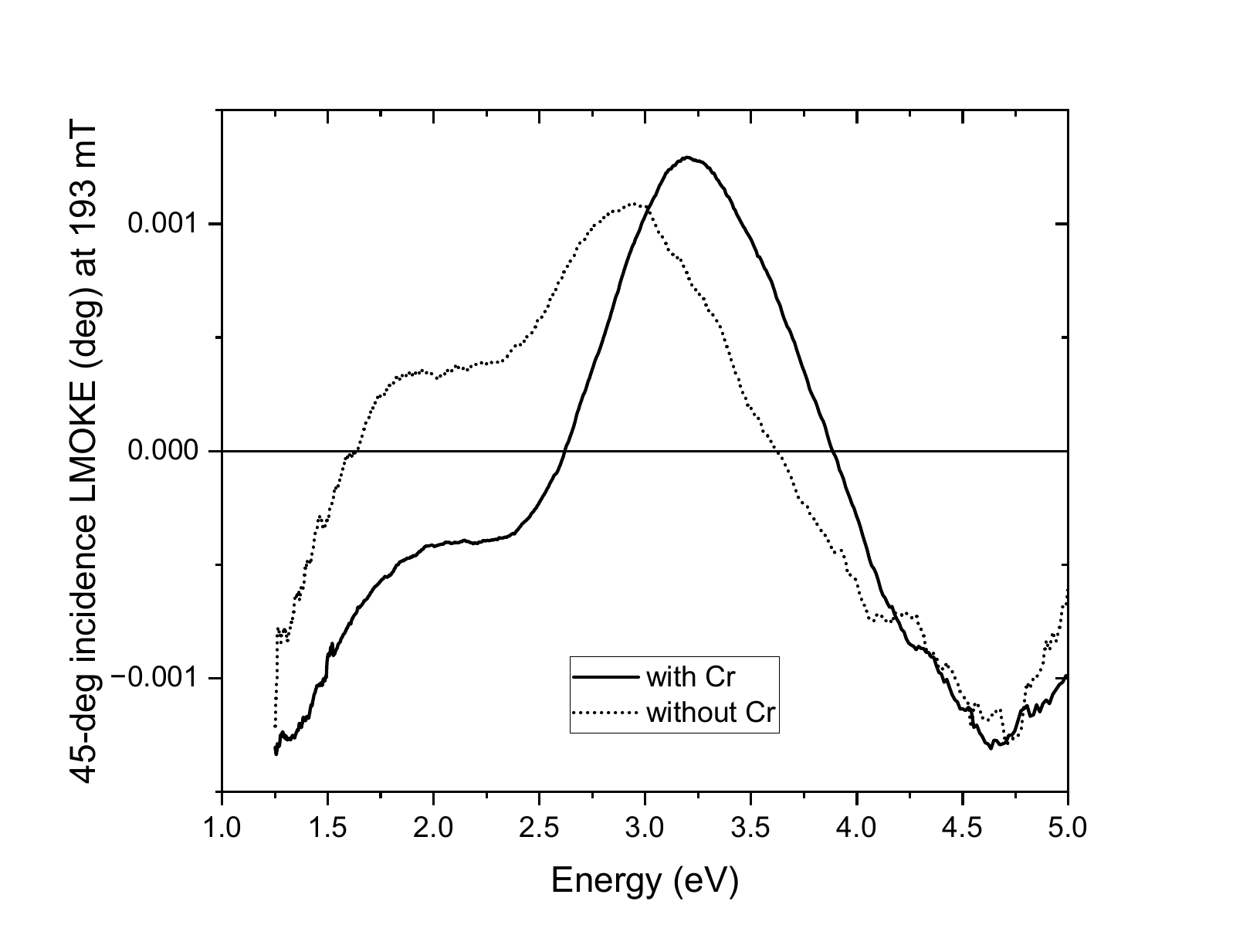}
    \caption{Influence of Cr buffer on 45-deg incidence LMOKE of 10 nm Ni-Mn-Ga film with \SI{193}{\milli\tesla} field applied along $\qty[100]$.}
    \label{fig:Cr_infl_LMOKE_AZ}
\end{figure}

\begin{figure}
     \centering
     \includegraphics[width=0.7\linewidth]{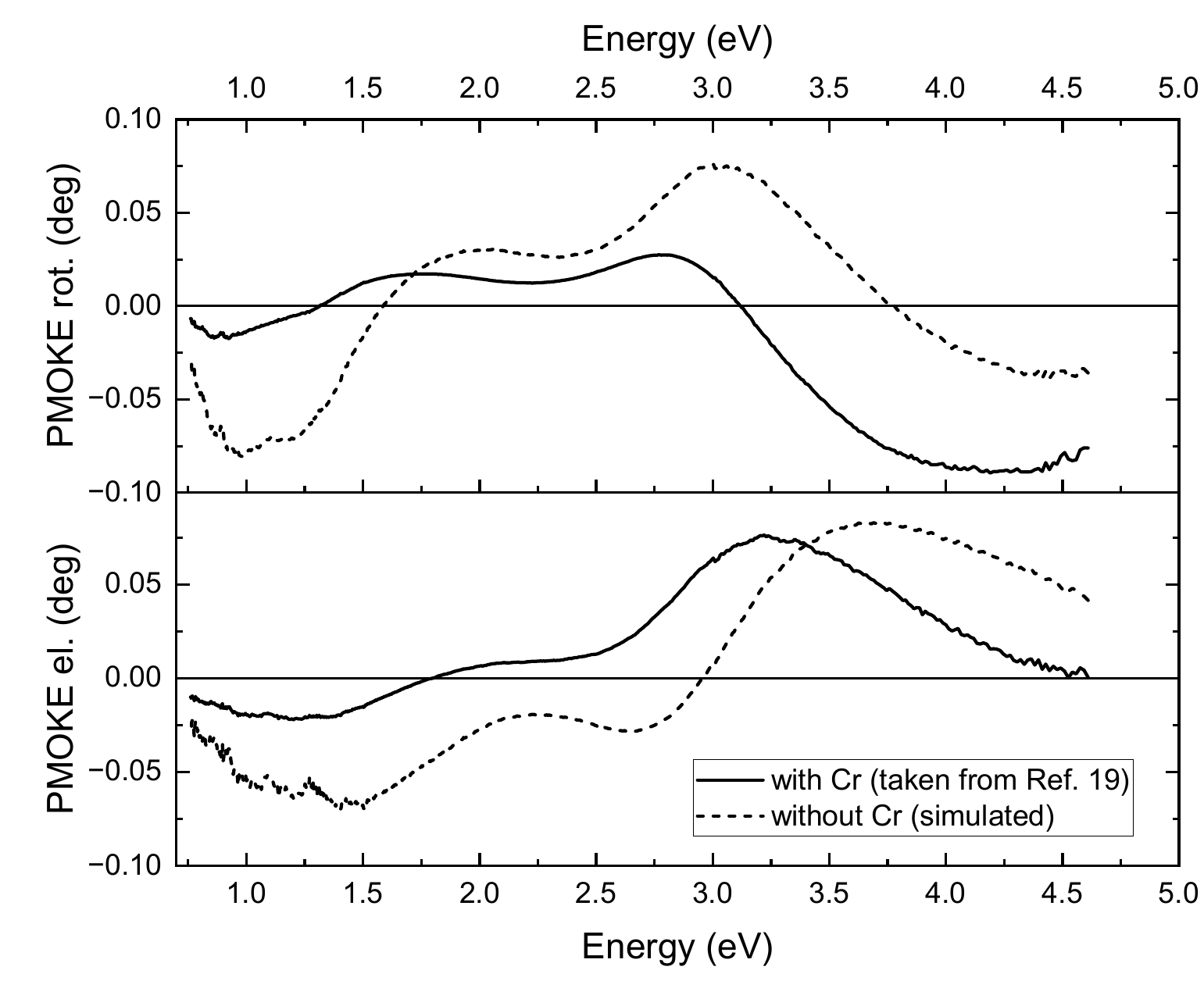}
     \caption{Normal incidence PMOKE rotation (top) and ellipticity (bottom) spectra of 16-nm thick Ni-Mn-Ga film on 28-nm thick Cr buffer (solid lines; data were taken from Ref.~\citenum{makes2026}) and a simulated PMOKE spectra (dashed lines) of the same film with the Cr buffer absent.}
     \label{fig:PMOKE_simulated_Cr_absence}
 \end{figure}

Cr-buffered Ni-Mn-Ga films \SIrange{8}{160}{\nm} thick show the same sequence of extrema in the PMOKE rotation, with negative extrema around \SIlist{1;3.8}{\eV} and two smaller positive ones around \SIlist{1.7;2.6}{\eV}, of which the former is higher in the martensitic and the latter in the austenitic films~\cite{makes2026}. Such a phase dependence was first observed in a bulk single crystal, in which the PMOKE rotation changes its sign between 2 and \SI{2.5}{\eV} upon the transformation between 10M martensite and austenite~\cite{beran2015}. In the films, however, the ratio of the two positive extrema does not depend on the phase alone, as the spectra also evolve with thickness. The films up to \SI{38}{\nm} thick are austenitic and compressed in-plane by the Cr buffer, with a tetragonal distortion $c/a>1$ that increases with thickness, and their PMOKE extrema change in amplitude and shift in energy by up to \SI{0.7}{\eV} with thickness. The \SIlist{80;160}{\nm} thick films are martensitic, but with a complex twinned microstructure in which the short axis of the martensitic cell lies in-plane and out-of-plane in roughly equal fractions. The film on Cr studied here ($c/a=1.021$) thus belongs to the same class of in-plane compressed austenite as the thinner films of Ref.~\cite{makes2026}, although it was grown from a different target, and its positive extremum at \SI{2.9}{\eV} and UV extremum at \SI{4.25}{\eV} lie close to those calculated for Cr-buffered films of similar thickness (solid lines in Fig.~\ref{fig:PMOKE_simulated_Cr_absence}). The film on bare MgO ($c/a=0.989$) is distorted in the opposite sense, which was not subject of the mentioned study, and its comparison with the twinned martensitic films, which contain both orientations of the short axis, is not straightforward.
The longitudinal MOKE (LMOKE) spectra of the \SI{10}{\nm} thick films differ between the two samples in a similar way (Fig.~\ref{fig:Cr_infl_LMOKE_AZ}). The LMOKE rotation peak near \SI{2}{\eV} has opposite signs in the two films, and the higher-energy peak lies at \SI{3.2}{\eV} in the film on Cr and at \SI{2.9}{\eV} in the film on bare MgO.

Since the optical penetration depth is comparable to the film thickness, the layer beneath the film also affects the measured spectra by optical means, through the reflection at the film/substrate interface. To estimate this effect, we simulated the removal of the Cr buffer from \SI{16}{nm} Ni-Mn-Ga film from Ref.~\citenum{makes2026} using the Yeh $4\times4$ matrix formalism~\cite{yeh1980}. The model consisted of a magnetized Ni-Mn-Ga layer, described by the diagonal ($\varepsilon_{xx}$) and off-diagonal ($\varepsilon_{xy}$) elements of its permittivity tensor, an isotropic Cr layer, and a semi-infinite MgO substrate, at normal incidence in the \SIrange{0.76}{4.6}{\eV} photon energy range. We used the permittivity tensors of the \SIlist{16}{\nm} Ni-Mn-Ga film with \SI{28}{\nm} thick Cr buffer from our previous study~\cite{makes2026}, whose $\varepsilon_{xy}$ was determined from PMOKE with the same layer model~\cite{makes2026}, together with the permittivities of a Cr layer and MgO obtained on calibration samples. Results of the simulations (Fig.~\ref{fig:PMOKE_simulated_Cr_absence}) suggest that removing the buffer increases the amplitudes and sign changes for some photon ranges, but the overall spectral shape is somewhat conserved. The optical effect of the buffer could explain the stronger low-energy rotation extremum of the film on bare MgO, but not its other spectral changes, which we therefore attribute mainly to the change of the electronic structure of the film caused by the opposite strain.

Near $E_\text{F}$, the tetragonal distortion acts mainly on the minority-spin Ni $3d$ states of $e_g$ symmetry, which in the tetragonal cell separate into $a_{1g}$ ($d_{3z^2-r^2}$) and $b_{1g}$ ($d_{x^2-y^2}$) sub-bands~\cite{obata2023}. GGA calculations describe the qualitative changes of these sub-bands with $c/a$. Ayuela et al.~\cite{ayuela2002} showed that the splitting is mainly due to Ni and proportional to the deformation, with little change of the Mn bands near $E_\text{F}$ for $c/a<1$, and that at their calculated minima, $c/a=0.94$ and $1.26$, the lower sub-band has $d_{x^2-y^2}$ character for $c/a<1$ and $d_{3z^2-r^2}$ character for $c/a>1$. The same order was found by Barman et al.~\cite{barman2005} at $c/a=0.97$ and by Obata et al.~\cite{obata2023} at $c/a=1.20$. The position of these sub-bands relative to $E_\text{F}$, however, is not reliable in GGA, which places the cubic peak about \SI{0.17}{\eV} below its position at $E_\text{F}$ obtained with QSGW~\cite{obata2023}. In the GGA calculations of Barman et al., the upper sub-band therefore reaches $E_\text{F}$ only at $c/a=0.96$ or $1.1$, whereas with the peak at $E_\text{F}$, as in QSGW, even the small splitting in our films is expected to shift the upper sub-band partly above $E_\text{F}$ and the lower one further below. The empty minority-spin Ni $e_g$ states just above $E_\text{F}$, which can act as final states of optical transitions, would then have $d_{3z^2-r^2}$ character in the film with $c/a=0.989$ and $d_{x^2-y^2}$ character in the film with $c/a=1.021$. A change of the unit-cell volume, which in the same GGA calculations shifts the minority-spin Ni peak relative to $E_\text{F}$, plays no role here, as the volumes of the two films are practically equal. The positive rotation peak at \SI{1.45}{\eV}, the most prominent feature of the film with $c/a<1$, lies in the region below \SI{2}{\eV}, where the magneto-optical activity of Ni-Mn-Ga has been tentatively attributed to transitions involving Ni $3d$ states~\cite{veis2014}. The spectral changes, however, extend over the whole measured range. Since the PMOKE spectra depend on both the diagonal and the off-diagonal permittivity, assigning individual spectral features to electronic transitions would require the permittivity tensors of both films and magneto-optical calculations for small distortions of both signs, which have not been reported.

In summary, \SIrange{10}{15}{\nm} thick epitaxial Ni-Mn-Ga films grown on bare MgO(001) and on a Cr buffer are tetragonally distorted in opposite directions, with $c/a=0.989$ and $1.021$, while their unit-cell volumes and room-temperature magnetization loops are practically the same. Despite these small distortions, their PMOKE spectra differ significantly over the whole \SIrange{0.76}{5}{\eV} range in the amplitude, sign, and energy of the spectral features. Our optical simulations showed that the absence of the Cr buffer explains the stronger low-energy rotation extremum of the film on MgO but not the other changes, which we attribute to the strain-induced change in the electronic structure of the film. These changes are consistent with the calculated splitting of the minority-spin Ni $e_g$ states, in which the $d_{x^2-y^2}$ and $d_{3z^2-r^2}$ sub-bands exchange their order when the sign of the distortion changes~\cite{ayuela2002}. Magneto-optical spectroscopy thus detects strain-induced changes of the electronic structure in films only about \SI{10}{\nm} thick, even where magnetometry shows no difference, and the choice of the buffer layer may be used to modify the magneto-optical response of such ultra-thin films. In the rotation between about 2.5 and \SI{3.3}{\eV} and in the ellipticity between about 3.4 and \SI{4.6}{\eV}, the two films show MOKE of opposite sign, so that the choice of the template can switch the sign of the magneto-optical response at a given photon energy, which may be of interest for magneto-optical devices.

\section*{Declaration of competing interest}

The authors declare that they have no known competing financial interests or personal relationships that could have appeared to influence the work reported in this paper.

\section*{Data availability}

The data that support the findings of
this study are openly available in
Zenodo, reference number \citenum{makes2026datajalcom}.

\section*{Declaration of generative AI and AI-assisted technologies in the manuscript preparation process}

During the preparation of this work, the authors used Claude Opus 5.5 (Anthropic) to improve the readability and structure of the text, to check the cited statements against the original publications, and for literature research. The authors reviewed and edited all output of this tool and take full responsibility for the content of the publication.

\section*{Acknowledgements}

{\sloppy This work was supported by Czech Science Foundation Grant No 24-11361S, OPJAK - Ferroic multifunctionalities (FerrMion) CZ.02.01.01/00/22\_008/0004591, and the Charles University grant SVV–2025–260836.\par}

\bibliographystyle{elsarticle-num-names}
\bibliography{references}

\end{document}